# Nanoscale quantum imaging of field-free deterministic switching of a chiral antiferromagnet

Jingcheng Zhou[1], Senlei Li[1], Chuangtang Wang[2], Hanshang Jin[3], Stelo Xu[3], Zelong Xiong[1], Carson Jacobsen[1], Kenji Watanabe[4], Takashi Taniguchi[5], Valentin Taufour[3], Liuyan Zhao[2], Hua Chen[6,7], Chunhui Rita Du[1,*], and Hailong Wang[1,*]

[1]School of Physics, Georgia Institute of Technology, Atlanta, Georgia 30332, USA
[2]Department of Physics, University of Michigan, Ann Arbor, Michigan 48109, USA
[3]Department of Physics and Astronomy, University of California, Davis, California 95616, USA
[4]Research Center for Electronic and Optical Materials, National Institute for Materials Science, Tsukuba 305-0044, Japan
[5]Research Center for Materials Nanoarchitectonics, National Institute for Materials Science, Tsukuba 305-0044, Japan
[6]Department of Physics, Colorado State University, Fort Collins, Colorado 80523, USA
[7]School of Materials Science and Engineering, Colorado State University, Fort Collins, Colorado 80523, USA

*Corresponding authors: cdu71@gatech.edu, hwang3021@gatech.edu

**Abstract**: Recently, unconventional spin-orbit torques (SOTs) with tunable spin generation open new pathways for designing novel magnetization control for cutting-edge spintronics innovations. A leading research thrust is to develop field-free deterministic magnetization switching for implementing scalable and energy favorable magnetic recording and storage applications, which have been demonstrated in conventional ferromagnetic and antiferromagnetic material systems. Here we extend this advanced magnetization control strategy to chiral antiferromagnet $Mn_3Sn$ using spin currents with out-of-plane canted polarization generated from low-symmetry van der Waals (vdW) material $WTe_2$. Numerical calculations suggest that damping-like SOT of spins injected perpendicular to the kagome plane of $Mn_3Sn$ serves as a driving force to rotate the chiral magnetic order, while the field-like SOT of spin currents with polarization parallel to the kagome plane provides the bipolar deterministicity to the magnetic switching. We further introduce scanning quantum microscopy to visualize nanoscale evolutions of $Mn_3Sn$ magnetic domains during the field-free switching process, corroborating the exceptionally large magnetic switching ratio up to 90%. Our results highlight the opportunities provided by hybrid SOT material platforms consisting of noncollinear antiferromagnets and low-symmetry vdW spin source materials for developing next-generation, transformative spintronic logic devices.

Harnessing spin-orbit torques (SOTs) to realize advanced magnetization control constitutes a key strategy of modern spintronic technologies[1,2]. Exploring innovative SOT material platforms with tailored, on-demand spin properties is a key step to fulfill this goal[1,3,4]. The family of emergent spin source materials, such as $WTe_2$[5–9], CuPt[10], and $TaIrTe_4$[11,12], with reduced crystal symmetry is directly relevant in this context thanks to their unique capability of generating spin currents with tunable polarization orientations. To date, out-of-plane spin-assisted field-free deterministic switching has been experimentally realized in a range of perpendicular magnet-based material systems, providing an excellent solution to address the longstanding technical issues related to solid-state scalability and energy consumption of conventional SOT technologies[6–9,11–15]. Despite the significant progress and technological benefits, to date, this novel magnetic control strategy was mainly implemented in the family of ferrimagnets and ferromagnets while the role of other emergent magnetic materials with unconventional spin orders remains peripheral.

Chiral antiferromagnets represented by $Mn_3X$ (X = Sn, Ge, Ga, Ir, Pt, Rh) compound with vanishingly small net magnetization, robust magneto-transport responses, and frustrated magnetic interaction naturally stand out as an appealing material candidate to develop advanced SOT operations[16–20]. The noncollinear antiferromagnetic order of $Mn_3X$ promises to deliver transformative magnetic recording and storage functionalities with improved stability, reproducibility, terahertz-scale information processing speeds, and ultra-high solid-state densities that outperform (ferri)ferromagnetic spintronic deivices[16,20,21]. Pioneering research efforts have been dedicated to studying SOT-driven control of chiral antiferromagnetic order[13,16,21–26], and state-of-the-art $Mn_3X$-based spintronic units such as magnetic tunnel junctions[27–29], antiferromagnetic spin-torque diode[30], and nano-oscillators[31] are underway.

Taking advantage of unconventional SOTs arising from low-symmetry van der Waals (vdW) material $WTe_2$[5], here we report field-free deterministic switching of a prototypical noncollinear antiferromagnet $Mn_3Sn$. Micromagnetic simulations reveal that spins injected perpendicular to the kagome plane of $Mn_3Sn$ serve as the driving force to rotate the chiral magnetic order, while spin currents with polarization parallel with the kagome plane play the role of an effective magnetic field to achieve bipolar magnetization control. We further utilized scanning quantum microscopy[32–40] to visualize nanoscale evolutions of $Mn_3Sn$ magnetic domains during the field-free deterministic switching process, corroborating the exceptionally large magnetic switching ratio up to 90%. The presented results suggest that the family of chiral antiferromagnets and vdW spin-sources with reduced crystal symmetry could provide an attractive SOT material combination for designing next-generation, multifunctional spin logic devices. Our study also highlights the advantages of quantum spin sensors in probing the microscopic magnetic structures and spin properties of noncollinear antiferromagnets.

We first review the material/device platform used for electrical SOT and scanning nitrogen-vacancy (NV) imaging measurements in the current study as shown in Fig. 1a. Exfoliated $WTe_2$ nanoflakes were mechanically transferred onto prepatterned $Mn_3Sn$ Hall devices with Au pads arranged for four-probe Hall voltage measurements. The top surface of the $Mn_3Sn$ sample was gently ion beam etched before stacking the $WTe_2$ flake to ensure a clean and sharp interfacial condition. Hexagonal boron nitride (hBN) encapsulation layers covering the Hall cross area were utilized to prevent sample degradation. The device preparation process was performed in a glove box filled with argon to minimize environmental effects (see Method Section and Supplementary Information Note 1 for details). Chiral antiferromagnet $Mn_3Sn$ features the inverse-triangular spin configuration with hexagonal $D0_{19}$ structure as shown in Fig. 1b[16,20,21]. The kagome lattice of Mn atoms shows the characteristic ABAB stacking geometry, and the spin orientations of the three sublattice Mn moments form an angle of ~120° with a chirality set by the Dzyaloshinskii–Moriya interaction[16]. Symmetry-allowed spin canting of individual Mn moment introduces a vanishingly small remnant magnetization in the magnetic easy plane (0001), and the six non-collinearly ordered Mn atoms on two neighboring kagome planes define a nonzero cluster magnetic octupole moment[16,20] or electronic chiralization[41]. In the current study, we used magnetron sputtering technique to deposit polycrystalline $Mn_3Sn$ films. The magnetic, electrical transport, and crystal properties of prepared $Mn_3Sn$ samples have been systematically evaluated as reported in Supplementary Information Note 1.

We next discuss the unconventional symmetry-enabled spin generation in spin source material $WTe_2$ in the proposed SOT system. Transition-metal dichalcogenide $WTe_2$ was theoretically predicted and recently demonstrated as a material candidate capable of generating out-of-plane spins due to its lack of lateral mirror symmetry (in respective to the *ac*-plane) as shown in Fig. 1c[5–9]. When an in-plane charge current flows along its high-symmetry crystallographic axis (*b*-axis), the directions of electric field, spin current, and spin polarization are mutually orthogonal as shown in Fig. 1d. In contrast, an electric charge current following along the low-symmetry crystallographic axis (*a*-axis) of $WTe_2$ is expected to generate spin currents with out-of-plane polarization in addition to in-plane polarized spin currents arising from the conventional spin Hall and/or Rashba-Edelstein effects (Fig. 1e)[5–9]. To determine the crystal symmetry and label the in-plane crystal axes of $WTe_2$, we performed rotational anisotropy second harmonic generation (RA-SHG) measurements. Because the *bc* mirror plane forbids SHG polarization along its normal direction, the measured SHG intensity drops to zero when the SHG polarization is set along the *a*-axis[42]. Figure 1f presents a representative polar plot of RA-SHG data taken on a ~35-nm-thick $WTe_2$ flake. Based on the indicated SHG polarization, the SHG extinction at 90° corresponds to the *b*-axis in the *bc* mirror plane and its normal 0° direction defines the *a*-axis.

Generation of both in-plane and out-of-plane polarized spins ($\sigma_y$ and $\sigma_z$) in $WTe_2$ provides new opportunities for engineering SOT-driven spin dynamics of chiral antiferromagnet $Mn_3Sn$. In the current study, we are mainly interested in geometry where the effective polarization of spin currents (generated by $WTe_2$) is canted away from the $Mn_3Sn$ kagome planes. Figures 2a and 2b show the schematic of damping-like and field-like SOT effective fields on the individual $Mn_3Sn$ sublattice moments under the proposed experimental condition. Invoking a classical SOT physical picture, the damping-like torque (with a sufficiently large magnitude) could overcome the local magnetic anisotropy and Gilbert damping to drive continuous rotation of the chiral magnetic order parameter of $Mn_3Sn$. The field-like torque arising from canted spins introduces the secondary tuning knob. The in-plane (relative to the kagome planes) component of field-like SOT effective fields naturally define two bipolar magnetic easy states (depending on the polarity of the applied electric currents) that the rotating chiral spin structure will finally settle into, enabling field-free deterministic magnetic switching of the chiral $Mn_3Sn$ spin order between two perpendicular magnetic states.

We now present our magneto-transport results to examine the proposed SOT switching picture. Figure 2c shows an optical microscopy image of a prepared $WTe_2$ (37)/$Mn_3Sn$ (50) device, where the numbers in parentheses indicate the thickness of each layer in nanometers. Figure 2d presents the anomalous Hall loop of the $Mn_3Sn$/$WTe_2$ device (device A) as a function of the out-of-plane magnetic field $B_{ext}$. The characteristic "negative" anomalous Hall resistance of $Mn_3Sn$ changes sign with the reversal of the external magnetic field, in agreement with previous studies[16,43,44]. The robust anomalous Hall effect of $Mn_3Sn$ provides a convenient way to detect its noncollinear antiferromagnetic order. Figures 2e shows electric current driven anomalous Hall loops of Mn3Sn/WTe2 (device A) without the assistance of an external magnetic field (see Methods for details). When the electrical write current pulse is applied along the low-symmetry crystallographic axis (*a*-axis), we observed robust field-free deterministic magnetic switching of Mn3Sn. The measured Hall voltage signals show positive and negative jumps at the critical write currents. Remarkably, the magnetic switching ratio of Mn3Sn is above 70% when normalized to the field induced switching measurements, which is significantly larger than that observed in polycrystalline Mn3Sn/heavy-metal SOT systems[16,44–46] (see Supplementary Information Note 2 for details). In contrast, when the electrical write current flows along the high-symmetry crystallographic axis (*b*-axis), the field-free magnetic switching signature disappears. Figure 2f presents a set of similar field-free deterministic magnetic switching results with the switching ratio up to 90% measured on device B, confirming the consistency of our results (see Supplementary Information Note 2 for details). Control experiments further demonstrate that self-induced magnetic switching and Joule heating effect could not play a major role in the observed field-free deterministic magnetic switching of Mn3Sn[24] (see Supplementary Information Notes 2 and 3 for details).

After showing the electrical SOT measurement results, next we perform numerical calculations based on a macrospin model to understand the observed field-free deterministic chiral antiferromagnetic switching behavior. Given the polycrystalline nature of the prepared Mn3Sn samples, variations of the measured anomalous Hall signals mainly result from field or current induced rotation of chiral spin structure with the kagome planes along the sample thickness direction (naturally defined as the *z*-axis). Here, we mainly focus on the situation that the electrical write current flows along the low-symmetry crystallographic axis (*a*-axis) of WTe2 and the kagome layer of Mn3Sn lies in the *ac*-plane of WTe2. Note that the *a*, *b*, and *c* crystallographic axis of WTe2 correspond to *x*, *y*, and *z* axis in the local coordinate frame as shown in Fig. 3a. In this case, the effective spin polarization ***σ*** considering the combined effects of reduced crystal symmetry and spin-orbit interaction lies in the *y-z* plane with a canting angle relative to the *y*-axis. Spin dynamics of individual sublattice magnetic moments ***m****$_i$* of Mn3Sn can be described by the coupled Landau-Lifshitz-Gilbert (LLG) equations as follows:[16,44]

$$\frac{\partial \boldsymbol{m}_i}{\partial t} = -\gamma \boldsymbol{m}_i \times \mathbf{B}_{\text{eff},i} + \alpha \boldsymbol{m}_i \times \frac{\partial \boldsymbol{m}_i}{\partial t} - \frac{\gamma \hbar \theta_{\text{SH}} J}{2eM_S d} \boldsymbol{m}_i \times (\boldsymbol{m}_i \times \boldsymbol{\sigma}) - \eta_{FL} \frac{\gamma \hbar \theta_{\text{SH}} J}{2eM_S d} \boldsymbol{m}_i \times \boldsymbol{\sigma} \quad (1)$$

where $\gamma$ is gyromagnetic ratio, $\hbar$ is the reduced Planck constant, $\theta_{\text{SH}}$ is the charge-to-spin conversion efficiency of WTe2, *e* is the electron charge, $M_s$ and $\boldsymbol{B}_{\text{eff}} = -(M_s)^{-1} \delta u / \delta \boldsymbol{m}_i$ are the magnetization and the effective magnetic field of each magnetic sublattice, *u* is the magnetic energy density, ***σ*** is the polarization of spin currents, *α* and *d* are the Gilbert damping constant and the thickness of the Mn3Sn film, $\eta_{FL}$ is the field-like torque efficiency, and *J* is the electric current density in WTe2. Note that we have considered both damping-like [$\boldsymbol{m}_i \times (\boldsymbol{m}_i \times \boldsymbol{\sigma})$] and field-like ($\boldsymbol{m}_i \times \boldsymbol{\sigma}$) SOT effects arising from the canted spins in our simulations and the canting angle of ***σ*** is set to be ~75° relative to *y*-axis. Substituting the relevant material parameters with $J = \pm 1.45 \times 10^{11}$ A/m$^2$ and solving the above equation numerically, Fig. 3b presents the time-dependent variation of the weak ferromagnetic moment (average magnetic moment of all sublattices) projected along *x*- and *z*-axis directions (see Supplementary Information Note 4 for details). One can see that the uncompensated magnetic order $M_z$ exhibits the deterministic "up-to-down" and "down-to-up" switching behavior under negative and positive electrical write current application, respectively, in agreement with our experimental results. When out-of-plane spins ($\sigma_z$) are absent in our simulations, the net spin polarization ($\sigma_y$) is perpendicular to the kagome planes of Mn$_3$Sn (Fig. 3c). In this case, the chiral antiferromagnetic order of Mn$_3$Sn shows continuous oscillations while the electric current is applied and the final equilibrium position of the magnetic order sensitively depends on the duration of the current pulse as illustrated in the simulation results presented in Fig. 3d (see Supplementary Information Note 4 for details).

Here, we further comment on the underlying mechanism of the field-free deterministic chiral antiferromagnetic control observed in Mn$_3$Sn/WTe$_2$ SOT system. For Mn$_3$Sn with a weak magnetic anisotropy in the (0001) kagome plane and a large effective easy-plane magnetic anisotropy, SOT-induced chiral spin rotation is perceived to be an energy-favorable way to control its noncollinear antiferromagnetic order. Damping-like SOT [$\boldsymbol{m}_i \times (\boldsymbol{m}_i \times \boldsymbol{\sigma})$] arising from spins injected perpendicular with the magnetic easy plane (0001) are identically applied to all the three individual sublattice moments with the same direction. With a sufficiently large spin injection, the chiral antiferromagnetic spin order of Mn$_3$Sn will start to rotate continuously. For conventional perpendicular-magnet/heavy-metal SOT systems, an external auxiliary magnetic field is required to break the time reversal symmetry in order to achieve deterministic magnetic switching. In the proposed Mn$_3$Sn/WTe$_2$ SOT system, the deterministic switching nature of chiral antiferromagnet Mn$_3$Sn results from the field-effect of out-of-plane spins ($\sigma_z$) parallel with the kagome plane. The field-like SOT ($\boldsymbol{m}_i \times \boldsymbol{\sigma}$) will modify the energy landscape of the local magnetic order of Mn$_3$Sn, setting two energy-favorable magnetic states in the easy plane of the chiral spin order depending on the polarity of the applied electrical write currents. With the assistance of the field-like SOT of out-of-plane canted spins, the chiral antiferromagnetic order can eventually settle down to one of the two predefined magnetic easy states when the electrical spin injection is turned off.

We now utilize scanning single quantum microscopy[32–40] to spatially visualize the field-free deterministic switching in Mn$_3$Sn/WTe$_2$. Our quantum sensing studies exploit the Zeeman effect of an NV single electron spin contained in a diamond cantilever to detect local magnetic stray fields emanating from the Mn$_3$Sn sample (see Methods and Supplementary Information Note 5 for details)[33,45]. The spatial resolution is primarily determined by the vertical distance between the NV spin sensor and the sample surface, which is set to be ~60 nm in the current study[45]. Figure 4i shows a current driven field-free magnetic switching curve measured in Mn$_3$Sn/WTe$_2$ (device A). At a series of points ("A" to "H") marked on the magnetic hysteresis loop, we performed scanning NV measurements after application of individual electrical write current pulses to image SOT-induced variations of local magnetic field patterns of Mn$_3$Sn. Figures 4a-4h present representative stray field maps taken at the corresponding points ("A" to "H") on the current-induced magnetic hysteresis loop (Fig. 4i). It was noticed that the multidomain feature is observed in all the magnetic states, which is attributed to the polycrystalline nature of the Mn$_3$Sn device. Microscopically, a polycrystalline Mn$_3$Sn sample consists of weakly coupled crystalline grains with different local magnetic orientations, resulting in spatially varying stray field patterns showing opposite signs between neighboring domains[45–47]. Starting from the initial magnetic state ("A"), the measured stray field map barely changes when the electrical write current is below the threshold value. When ramping the write current above the critical value, local variations of the measured stray field start to emerge (Fig. 4b), leading to gradual reversal of the polarity of field domains (Fig. 4c). A certain amount of irreversible and inhomogeneous magnetic patterns is also observed between the two oppositely polarized magnetic states "A" and "D", which is potentially related to spatially dependent partial switching of the Mn$_3$Sn magnetization and/or the polycrystalline sample nature where some Mn$_3$Sn crystal grains with certain types of local magnetic orientations are not switchable by canted spins from WTe$_2$. When inverting the write current into the negative regime, the stray field domains gradually evolve back to the original state ("A") as shown in Fig. 4h. The presented scanning NV measurements image the nanoscale evolutions of polycrystalline Mn$_3$Sn magnetic domains during the field-free deterministic switching process, revealing the selective control of weakly coupled magnetic grains in prepared devices. We further corroborate the deterministic nature of the presented nonvolatile chiral antiferromagnetic control. Figure 5a shows the "step-like" variations of anomalous Hall signals of the Mn$_3$Sn/WTe$_2$ device in response to a train of positive and negative current pulses, $I = \pm 36$ mA, applied along the *a*-axis of WTe$_2$. Robust deterministic magnetization switching feature is highlighted in a series of scanning NV images recorded after individual current pulse applications as presented Figs. 5b-5e. It is evident that the positive and negative electric current pulses reliably switch the local Mn$_3$Sn domains between two deterministic magnetic states, in agreement with the "step-like" variations of anomalous Hall signals shown in Fig. 5a.

In summary, we have observed field-free deterministic switching of chiral antiferromagnet Mn$_3$Sn using spins with out-of-plane canted polarization generated from low-symmetry vdW material WTe$_2$. Micromagnetic simulations reveal that damping-like torque from spins injected perpendicular to the kagome planes serves as the driving force to rotate the chiral antiferromagnetic order of Mn$_3$Sn, while the field-like torque from spins parallel with the kagome plane provides the bipolar deterministicity of the magnetic switching. Taking advantage of scanning quantum microscopy, we further visualize the microscopic magnetic domain motions of Mn$_3$Sn during the field-free magnetic switching process. Our study provides an appealing SOT material platform built on chiral antiferromagnets and low-symmetry vdW quantum materials to develop advanced magnetic switching strategies for innovative spin information recording and storage applications. Elimination of external auxiliary magnetic fields in chiral antiferromagnet-based deterministic SOT switching experiments further paves the way for designing state-of-the-art spintronic logic devices with improved density, scalability, and processing speeds for practical applications[1–4].

**Acknowledgements**. The authors are grateful to Zhaorong Gu, Hanyi Lu, and Mengqi Huang for help on material preparations/characterizations and micromagnetic simulations. This work was primarily supported by the U.S. Department of Energy (DOE), Office of Science, Basic Energy Sciences (BES), under award No. DE-SC0024870. J. Z. acknowledges the support from the U.S. National Science Foundation under award No. DMR-2342569. Z. X. acknowledges the support from the U.S. National Science Foundation under award No. ECCS-2445826. L. Z. acknowledges the support from the U.S. Department of Energy (DOE), Office of Science, Basic Energy Science (BES), under award No. DE-SC0024145. The synthesis of WTe$_2$ crystals was supported by the Cahill Research Fund. H.C. acknowledges support from the National Science Foundation Grant No. DMR-1945023 and DMR-2414749.

**Competing interests**
The authors declare no competing interests.

**References**

1. Manchon, A. *et al.* Current-induced spin-orbit torques in ferromagnetic and antiferromagnetic systems. *Rev. Mod. Phys.* **91**, 35004 (2019).
2. Liu, L. *et al.* Spin-torque switching with the giant spin hall effect of tantalum. *Science* **336**, 555–558 (2012).
3. Yang, H. *et al.* Two-dimensional materials prospects for non-volatile spintronic memories. *Nature* **606**, 663–673 (2022).
4. Rimmler, B. H., Pal, B. & Parkin, S. S. P. Non-collinear antiferromagnetic spintronics. *Nat. Rev. Mater.* **10**, 109–127 (2025).
5. MacNeill, D. *et al.* Control of spin–orbit torques through crystal symmetry in $WTe_2$/ferromagnet bilayers. *Nat. Phys.* **13**, 300–305 (2017).
6. Kao, I.-H. *et al.* Deterministic switching of a perpendicularly polarized magnet using unconventional spin–orbit torques in $WTe_2$. *Nat. Mater.* **21**, 1029–1034 (2022).
7. Shin, I. *et al.* Spin–orbit torque switching in an all-van der Waals heterostructure. *Adv. Mater.* **34**, 2101730 (2022).
8. Zhang, Y. *et al.* Robust field-free switching using large unconventional spin-orbit torque in an all-van der Waals heterostructure. *Adv. Mater.* **36**, 2406464 (2024).
9. Kajale, S. N., Nguyen, T., Hung, N. T., Li, M. & Sarkar, D. Field-free deterministic switching of all–van der Waals spin-orbit torque system above room temperature. *Sci. Adv.* **10**, eadk8669 (2025).
10. Liu, L. *et al.* Symmetry-dependent field-free switching of perpendicular magnetization. *Nat. Nanotechnol.* **16**, 277–282 (2021).
11. Liu, Y. *et al.* Field-free switching of perpendicular magnetization at room temperature using out-of-plane spins from $TaIrTe_4$. *Nat. Electron.* **6**, 732–738 (2023).
12. Zhang, Y. *et al.* Room temperature field-free switching of perpendicular magnetization through spin-orbit torque originating from low-symmetry type II Weyl semimetal. *Sci. Adv.* **9**, eadg9819 (2025).
13. Zheng, Z. *et al.* All-electrical perpendicular switching of chiral antiferromagnetic order. *Nat. Mater.* (2025) doi:10.1038/s41563-025-02228-4.
14. Karube, S. *et al.* Observation of spin-splitter torque in collinear antiferromagnetic $RuO_2$. *Phys. Rev. Lett.* **129**, 137201 (2022).
15. Wu, Y. *et al.* Field-free spin-orbit switching of canted magnetization in Pt/Co/Ru/$RuO_2$(101) multilayers. *Appl. Phys. Lett.* **126**, 012401(2024).
16. Tsai, H. *et al.* Electrical manipulation of a topological antiferromagnetic state. *Nature* **580**, 608–613 (2020).
17. Nayak, A. K. *et al.* Large anomalous Hall effect driven by a nonvanishing Berry curvature in the noncolinear antiferromagnet $Mn_3Ge$. *Sci. Adv.* **2**, e1501870 (2025).
18. Cao, C. *et al.* Anomalous spin current anisotropy in a noncollinear antiferromagnet. *Nat. Commun.* **14**, 5873 (2023).
19. Jeon, K.-R. *et al.* Long-range supercurrents through a chiral non-collinear antiferromagnet in lateral Josephson junctions. *Nat. Mater.* **20**, 1358–1363 (2021).
20. Chen, H., Niu, Q. & MacDonald, A. H. Anomalous Hall effect arising from noncollinear antiferromagnetism. *Phys. Rev. Lett.* **112**, 17205 (2014).
21. Han, J., Cheng, R., Liu, L., Ohno, H. & Fukami, S. Coherent antiferromagnetic spintronics. *Nat. Mater.* **22**, 684–695 (2023).

22. Krishnaswamy, G. K. *et al.* Time-dependent multistate switching of topological antiferromagnetic order in $Mn_3Sn$. *Phys. Rev. Appl.* **18**, 24064 (2022).
23. Pal, B. *et al.* Setting of the magnetic structure of chiral kagome antiferromagnets by a seeded spin-orbit torque. *Sci. Adv.* **8**, eabo5930 (2025).
24. Xie, H. *et al.* Magnetization switching in polycrystalline $Mn_3Sn$ thin film induced by self-generated spin-polarized current. *Nat. Commun.* **13**, 5744 (2022).
25. Higo, T. *et al.* Perpendicular full switching of chiral antiferromagnetic order by current. *Nature* **607**, 474–479 (2022).
26. Yoon, J.-Y. *et al.* Handedness anomaly in a non-collinear antiferromagnet under spin–orbit torque. *Nat. Mater.* **22**, 1106–1113 (2023).
27. Chen, X. *et al.* Octupole-driven magnetoresistance in an antiferromagnetic tunnel junction. *Nature* **613**, 490–495 (2023).
28. Chou, C.-T. *et al.* Large Spin Polarization from symmetry-breaking Antiferromagnets in Antiferromagnetic Tunnel Junctions. *Nat. Commun.* **15**, 7840 (2024).
29. Qin, P. *et al.* Room-temperature magnetoresistance in an all-antiferromagnetic tunnel junction. *Nature* **613**, 485–489 (2023).
30. Sakamoto, S. *et al.* Antiferromagnetic spin-torque diode effect in a kagome Weyl semimetal. *Nat. Nanotechnol.* **20**, 216–221 (2025).
31. Shukla, A. & Rakheja, S. Spin-torque-driven terahertz auto-oscillations in noncollinear coplanar antiferromagnets. *Phys. Rev. Appl.* **17**, 34037 (2022).
32. Casola, F., van der Sar, T. & Yacoby, A. Probing condensed matter physics with magnetometry based on nitrogen-vacancy centres in diamond. *Nat. Rev. Mater.* **3**, 17088 (2018).
33. Li, S. *et al.* Observation of stacking engineered magnetic phase transitions within moiré supercells of twisted van der Waals magnets. *Nat. Commun.* **15**, 5712 (2024).
34. Song, T. *et al.* Direct visualization of magnetic domains and moiré magnetism in twisted 2D magnets. *Science* **374**, 1140–1144 (2021).
35. Thiel, L. *et al.* Quantitative nanoscale vortex imaging using a cryogenic quantum magnetometer. *Nat. Nanotechnol.* **11**, 677–681 (2016).
36. Pelliccione, M. *et al.* Scanned probe imaging of nanoscale magnetism at cryogenic temperatures with a single-spin quantum sensor. *Nat. Nanotechnol.* **11**, 700–705 (2016).
37. Finco, A. *et al.* Imaging non-collinear antiferromagnetic textures via single spin relaxometry. *Nat. Commun.* **12**, 767 (2021).
38. Monge, R. *et al.* Spin dynamics of a solid-state qubit in proximity to a superconductor. *Nano Lett.* **23**, 422–428 (2023).
39. Palm, M. L. *et al.* Observation of current whirlpools in graphene at room temperature. *Science* **384**, 465–469 (2024).
40. Guo, Q. *et al.* Current-induced switching of thin film $\alpha$-$Fe_2O_3$ imaged using a scanning single-spin microscope. *Phys. Rev. Mater.* **7**, 64402 (2023).
41. Chen, H. Electronic chiralization as an indicator of the anomalous Hall effect in unconventional magnetic systems. *Phys. Rev. B* **106**, 24421 (2022).
42. Drueke, E., Yang, J. & Zhao, L. Observation of strong and anisotropic nonlinear optical effects through polarization-resolved optical spectroscopy in the type-II Weyl semimetal $T_d$-$WTe_2$. *Phys. Rev. B* **104**, 64304 (2021).
43. Nakatsuji, S., Kiyohara, N. & Higo, T. Large anomalous Hall effect in a non-collinear antiferromagnet at room temperature. *Nature* **527**, 212–215 (2015).

44. Takeuchi, Y. *et al.* Chiral-spin rotation of non-collinear antiferromagnet by spin–orbit torque. *Nat. Mater.* **20,** 1364–1370 (2021).
45. Li, S. *et al.* Nanoscale Magnetic Domains in polycrystalline $Mn_3Sn$ films imaged by a scanning single-spin magnetometer. *Nano Lett.* **23**, 5326–5333 (2023).
46. Yan, G. Q. *et al.* Quantum sensing and imaging of spin–orbit-torque-driven spin dynamics in the non-collinear antiferromagnet $Mn_3Sn$. *Adv. Mater.* **34**, 2200327 (2022).
47. Tsukamoto, M. *et al.* Observation of chiral domain walls in an octupole-ordered antiferromagnet. Preprint at https://arxiv.org/abs/2410.23607 (2024).
48. Higo, T. et al. Anomalous Hall effect in thin films of the Weyl antiferromagnet $Mn_3Sn$. *Appl. Phys. Lett.* **113**, 202402 (2018).
49. You, Y. *et al.* Anomalous Hall effect–like behavior with in-plane magnetic field in noncollinear antiferromagnetic $Mn_3Sn$ films. *Adv. Electron. Mater.* **5**, 1800818 (2019).
50. Canfield, P. C. & and Fisk, Z. Growth of single crystals from metallic fluxes. *Philos. Mag. B* **65**, 1117–1123 (1992).
51. Canfield, P. C., Tai, K., Udhara S., K. & and Jo, N. H. Use of frit-disc crucibles for routine and exploratory solution growth of single crystalline samples. *Philos. Mag.* **96**, 84–92 (2016).
52. Canfield, P. C. & Slade, T. J. Use of frit-disc crucible sets to make solution growth more quantitative and versatile. *Z. Anorg. Allg. Chem.* **651**, e202500007 (2025).
53. Ali, M. N. *et al.* Correlation of crystal quality and extreme magnetoresistance of $WTe_2$. *Europhys. Lett.* **110**, 67002 (2015).
54. Zhang, X. *et al.* Visualizing field-free deterministic magnetic switching of all-van der Waals spin-orbit torque system using spin ensembles in hexagonal boron nitride. Preprint at https://arxiv.org/abs/2502.04561 (2025).

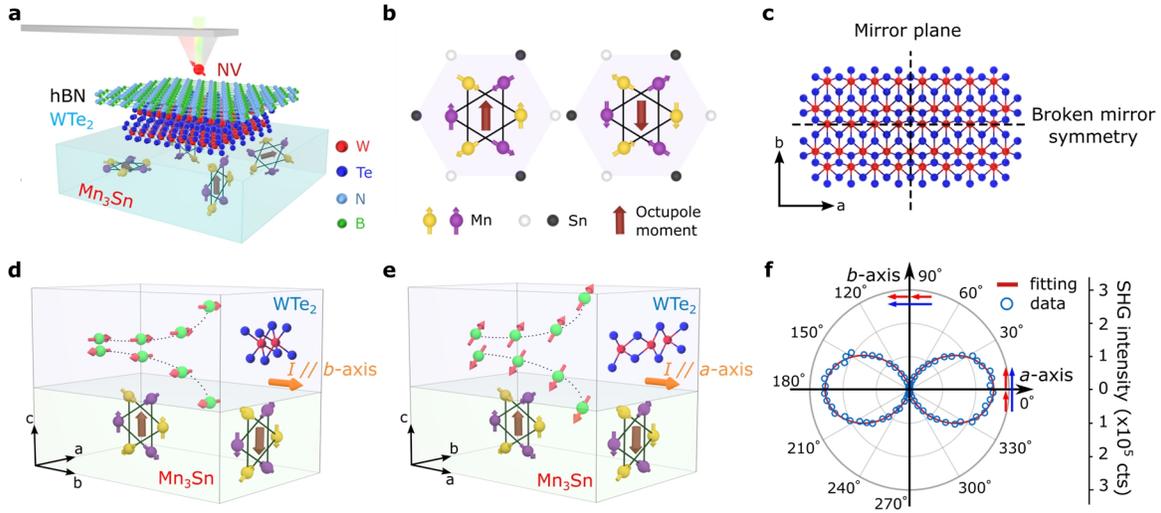

**Figure 1. Device structures and material properties. a** Schematic of the $Mn_3Sn/WTe_2/hBN$ material platform for SOT switching and scanning quantum imaging measurements. **b** Schematic of the kagome lattices of $Mn_3Sn$ hosting inverse triangular spin configurations. **c** Atomic arrangement of $WTe_2$ viewed along its *c*-axis showing a lack of crystal mirror symmetry in respective to the *ac*-plane. **d, e** Generation of in-plane polarized and out-of-plane canted polarized spin currents when an electric current flows along the high-symmetry and low-symmetry crystallographic axes of $WTe_2$. **f** RA SHG polar plot taken in the parallel linear polarization channel showing the lack of mirror in the *ac*-plane and the presence of mirror in the *bc*-plane. The solid line is the fitting of the functional form of the electric-dipole SHG response under the *m* point group with *bc*-mirror plane. Red and blue arrows denote the polarization direction of the fundamental and SHG light, respectively.

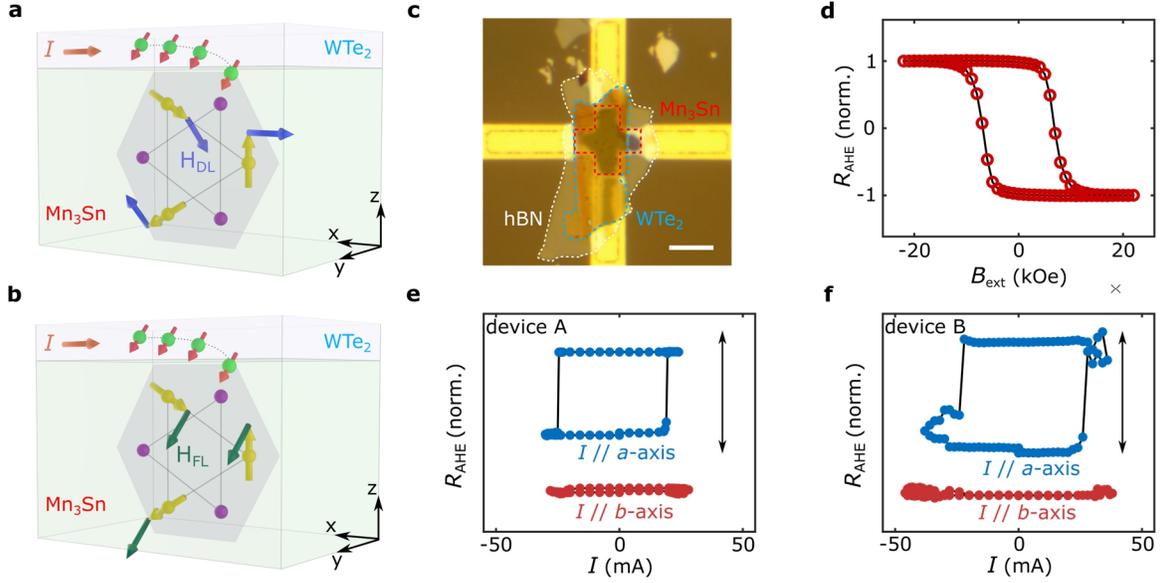

**Figure 2. Field-free deterministic switching of chiral antiferromagnet Mn₃Sn. a, b** Schematic of the damping-like and field-like SOT effective fields ($H_{DL}$ and $H_{FL}$) on individual Mn₃Sn sublattice moments. The Mn₃Sn kagome plane lies in *xz*-plane, and electric current flows along the negative *x*-axis direction with an effective spin polarization canted away from the kagome plane. **c** Optical microscopy image of a prepared Mn₃Sn/WTe₂/hBN device. The boundaries of Mn₃Sn Hall cross, WTe₂ and hBN encapsulation flakes are outlined by red (Mn₃Sn), blue (WTe₂), and white (hBN) dashed lines. Au electrodes are patterned for electrical SOT measurements. The scale bar is 20 μm. **d** Anomalous Hall loop of the Mn₃Sn/WTe₂ SOT device. **e, f** Normalized anomalous Hall resistance ($R_{AHE}$) of Mn₃Sn measured as a function of electrical write current pulse (*I*) for devices A and B, respectively. External auxiliary magnetic fields were absent in the presented SOT measurements. Field-free deterministic chiral antiferromagnetic switching is observed when *I* is applied along the *a*-axis of WTe₂. The black arrow shows the anomalous Hall difference between $m_z = -1$ and $m_z = +1$.

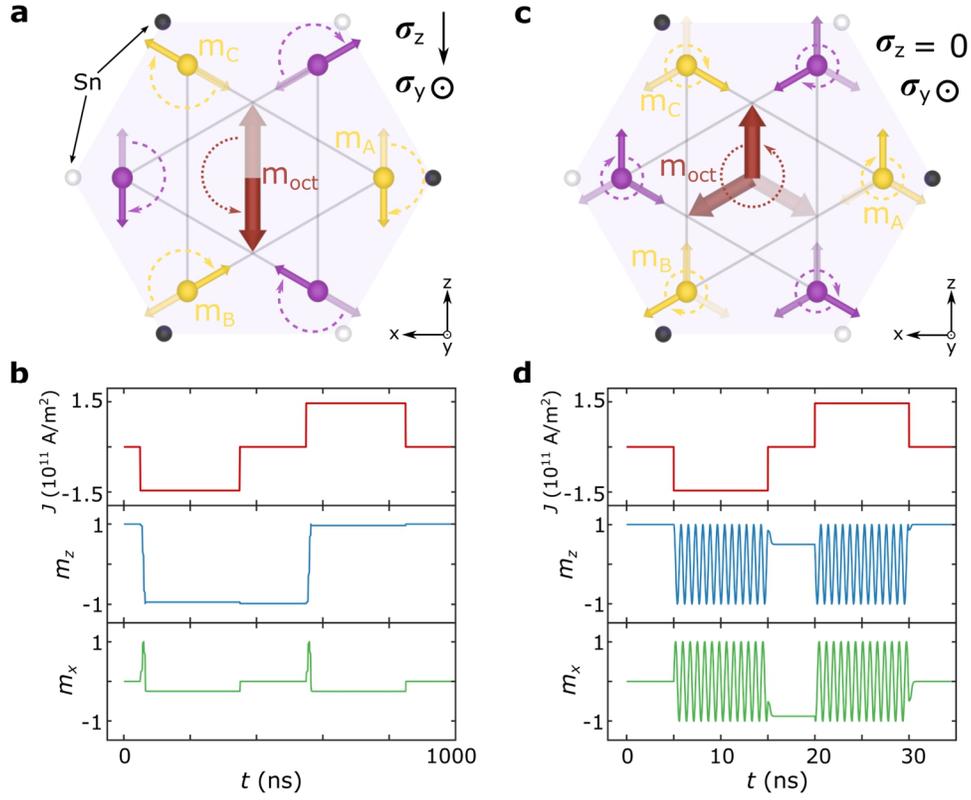

**Figure 3. Micromagnetic simulations of field-free deterministic switching of chiral antiferromagnetic order. a, b** Schematic illustration and micromagnetic simulations of field-free deterministic magnetic switching of Mn$_3$Sn driven by combined damping-like and field-like SOTs of canted spins. Spin polarization is in *yz*-plane with a canting angle of ~75° relative to the *y*-axis, and the Mn$_3$Sn kagome plane lies in *xz*-plane. **c, d** Schematic and simulations of continuous chiral spin rotation of Mn$_3$Sn driven by SOTs from spins perpendicular to the kagome plane. Spin polarization is along the *y*-axis direction and the Mn$_3$Sn kagome plane lies in *xz*-plane.

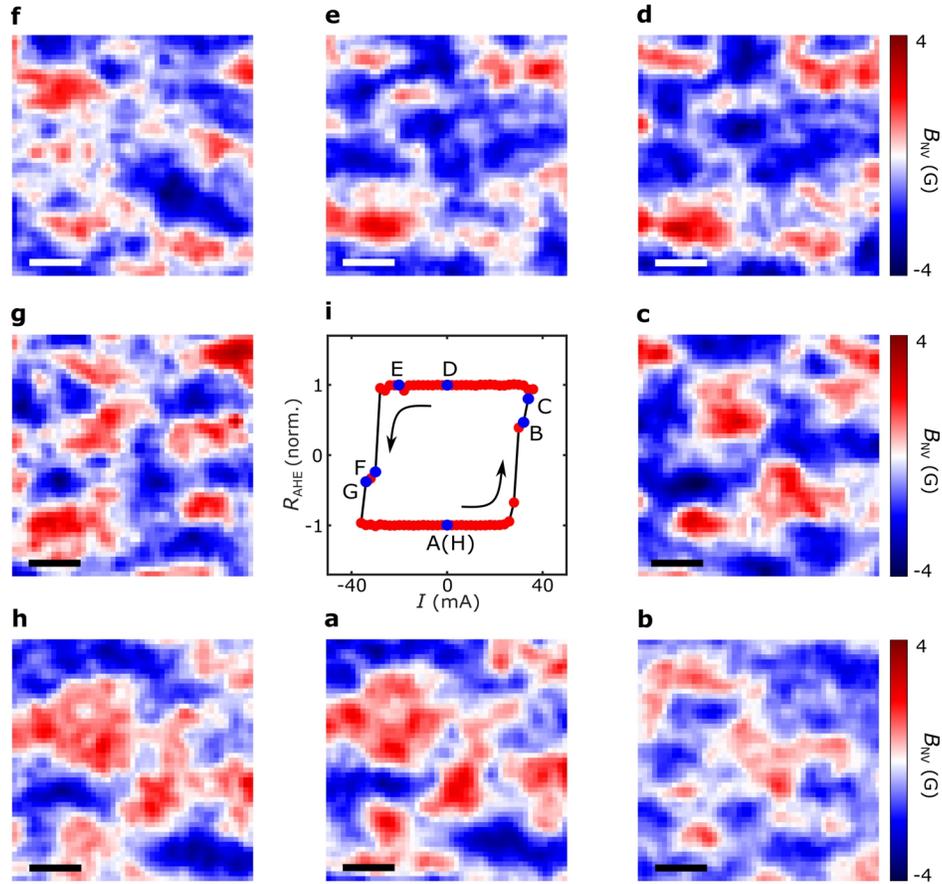

**Figure 4. Scanning quantum imaging of field-free deterministic magnetic switching of Mn₃Sn.**
**a-h** Scanning NV imaging of nanoscale evolutions of Mn₃Sn chiral antiferromagnetic domains during the field-free deterministic magnetic switching process. The scale bar is 400 nm. **i** Anomalous Hall resistance of Mn₃Sn measured as a function of electrical write current pulse *I* in absence of an external magnetic field. The arrows indicate that *I* was swept from zero following the counterclockwise direction around the hysteresis loop and finally returned to the starting point. Scanning NV imaging measurements presented in Figs. 4a-4h were performed at the corresponding points from "A" to "H" marked on the field-free deterministic switching loop (Fig. 4i).

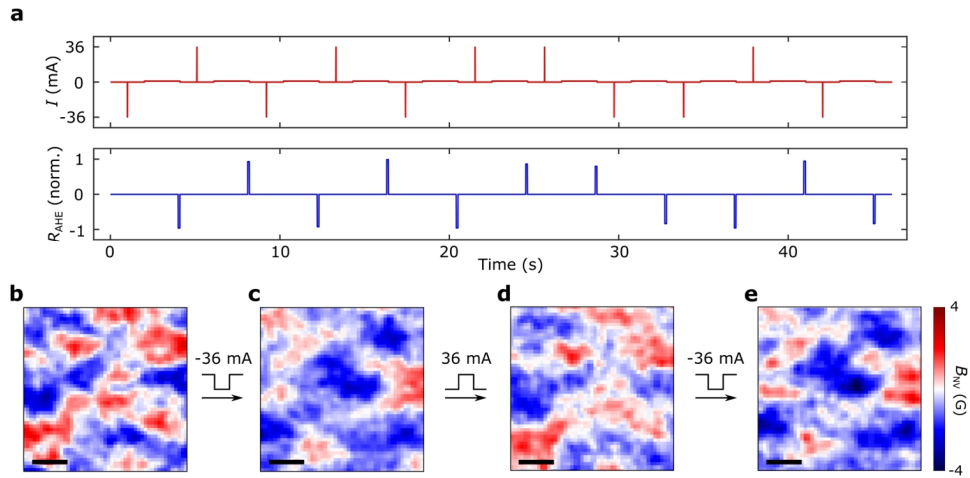

**Figure 5. Visualizing the deterministic nature of field-free chiral antiferromagnetic control.**
**a** Deterministic field-free chiral antiferromagnetic switching of Mn$_3$Sn using a train of positive and negative electric current pulses ($I$) with a magnitude of 36 mA applied along the low-symmetry $a$-axis of WTe$_2$. **b-e** Variations of magnetic stray field patterns of the Mn$_3$Sn/WTe$_2$ device after individual electric current pulse applications along the $a$-axis of WTe$_2$ for $I = \pm 36$ mA. The scale bar is 400 nm in all figures.